\documentclass[PAPER, REPORT=false, coverpage=false, atlasdraft=false, cernpreprint, USenglish, texmf, orcidlogo]{atlasdoc}
\usepackage[lineno=true,biblatex=true]{atlaspackage}
\newenvironment{DIFnomarkup}{}{}
\usepackage{atlasbiblatex}

\usepackage{atlasphysics}
\graphicspath{{logos/}{figures/}}

\usepackage{ANA-EXOT-2025-05-PAPER-defs}

\AtlasTitle{
Search for heavy Majorana neutrinos in vector boson scattering with $\tau$-lepton final states with the ATLAS detector
}

\AtlasAbstract{%

A search for heavy Majorana neutrinos coupling to third-generation leptons in vector boson scattering at the LHC is presented.
The search is performed in the final state with a same-sign $\tau\tau$, $e\tau$ or $\mu\tau$ pair using data of proton--proton collisions
at $\sqrt{s}$ = 13~TeV, recorded with the ATLAS detector and corresponding to an integrated luminosity of 140~$\text{fb}^{-1}$. No significant deviation from the Standard Model background is observed.
Constraints on the squared mixing matrix element between the $\tau$-lepton and the heavy Majorana neutrino ($|V_{\tau N}|^2$) are derived for heavy Majorana neutrino masses between 92~GeV and 6.5~TeV.
Additionally, two-dimensional constraints on ($|V_{eN}|^{2}, |V_{\tau N}|^{2}$) and ($|V_{\mu N}|^{2}, |V_{\tau N}|^{2}$) are set.
These results improve the limits on heavy Majorana neutrinos coupling to third-generation leptons for heavy Majorana neutrino masses above 1.0~TeV, and extend the exclusion limits to an unexplored mass range above 1.5~TeV.
}

\AtlasRefCode{EXOT-2025-05}

\PreprintIdNumber{CERN-EP-2026-221}

\AtlasJournal{PRL}
\AtlasCoverEgroupAnalysisTeam{atlas-exot-2025-05-analysis-team@cern.ch}

\hypersetup{pdftitle={ATLAS document},pdfauthor={The ATLAS Collaboration}}

\begin{document}
\maketitle

%

%
%

%
%
%
%

The observation of neutrino oscillations~\cite{Super-Kamiokande:2005mbp,SNO:2002tuh} has conclusively demonstrated that neutrinos have non-zero masses,
while stringent bounds from cosmological observations~\cite{Planck:2018vyg,eBOSS:2020yzd} and direct experimental measurements,
such as tritium $\beta$-decay kinematics~\cite{KATRIN:2024cdt}, indicate that neutrino masses are small.

The small but finite neutrino masses are generally assumed to be generated by a mechanism beyond the Standard Model (SM).
A natural approach can be realized in the context of ``seesaw'' models~\cite{Minkowski:1977sc,Mohapatra:1979ia,Schechter:1980gr},
assuming the existence of heavy states, such as a heavy Majorana neutrino ($N$).
In the context of the type-I seesaw mechanism~\cite{Atre:2009rg,Deppisch:2015qwa,Cai:2017jrq,Cai:2017mow}, the small neutrino masses $m_{\nu}$ are explained
by introducing a new heavy Majorana neutrino which scales $m_{\nu}$ to the level of $\mathcal{O} (\upsilon^2 / m_N)$,
where $\upsilon$ is the vacuum expectation value of the Higgs field and $m_{N}$ is the mass of the heavy Majorana neutrino.

The Majorana nature of neutrinos can be probed by searching for neutrinoless double beta decays of certain isotopes, which are characterized
by the appearance of two same-sign (SS) electrons with an absence of neutrinos in the final state~\cite{Dolinski:2019nrj}.
Since muons and $\tau$-leptons are much heavier than electrons, SS final states involving these heavier leptons are kinematically forbidden in low-energy nuclear experiments
and can only be studied at colliders~\cite{Fuks:2020zbm}.

Searches for heavy Majorana neutrinos in the $W^{\pm}W^{\pm}$ vector boson scattering (VBS) topology using pairs of same-sign electrons or muons in the final state at the Large Hadron Collider (LHC) were reported by the ATLAS~\cite{EXOT-2023-16,EXOT-2020-06} and CMS~\cite{CMS-EXO-21-003} Collaborations.
These searches probe higher heavy Majorana neutrino masses than searches via resonant $W$-boson production reported by the ATLAS~\cite{EXOT-2012-24, TOPQ-2019-31, EXOT-2019-35, EXOT-2017-26, EXOT-2022-12},
CMS~\cite{CMS-EXO-17-028,CMS-EXO-22-011,CMS-EXO-21-010,CMS-EXO-20-009,CMS-EXO-22-017, CMS-EXO-21-013,CMS-EXO-22-019} and LHCb~\cite{PhysRevLett.112131802, LHCb2026HNL} Collaborations, since resonant $W$-boson production becomes kinematically inaccessible beyond the \TeV~scale~\cite{Fuks:2020att}.

\begin{figure}[htbp]
\centering
\includegraphics[width=0.30\textwidth]{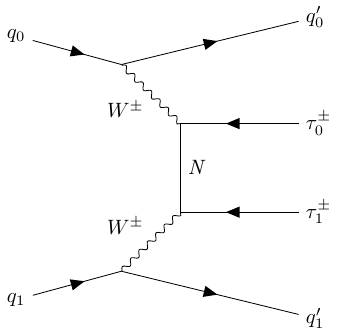}
\caption{Representative Feynman diagram of same-sign $\tau\tau$ production in vector boson scattering mediated by a heavy Majorana neutrino $N$.}
\label{fig:feyn_sig}
\end{figure}

This Letter presents the search for heavy Majorana neutrinos using a pair of same-sign $\tau$-leptons in the VBS topology at the LHC, as shown in Figure~\ref{fig:feyn_sig}, for the first time.
Both $\TauLepHad$ and $\HadHad$ decay channels are considered, where $\taulep$ denotes the decay of the $\tau$-lepton into two neutrinos and either an electron or a muon,
and $\tauhad$ denotes the decay into a $\tau$ neutrino and hadrons. In addition, the same-charge $e\tau$ and $\mu\tau$ final states are analyzed, where one of the $\tau$-leptons in the process shown in Figure~\ref{fig:feyn_sig} is replaced by a prompt electron or muon, respectively.
This search provides sensitivity to heavy Majorana neutrinos with $\mN > 1.0~\TeV$, extending the reach beyond previous searches and probing an unexplored mass range up to $\mN = 6.5~\TeV$.
Furthermore, simultaneous constraints on $(|V_{eN}|^2, |V_{\tau N}|^2)$ and $(|V_{\mu N}|^2, |V_{\tau N}|^2)$ are derived.

%
%
%
%


%

%
The ATLAS experiment at the LHC is a multipurpose particle detector with a forward--backward symmetric cylindrical geometry,
detailed in Refs.~\cite{PERF-2007-01}. It consists of an inner tracking detector (ID) surrounded by a thin superconducting solenoid,
electromagnetic and hadronic calorimeters, and a muon spectrometer.
The $pp$ collision data are recorded with the ATLAS detector totaling \SI{140}{\fb^{-1}} at $\sqrt{s}=13~\TeV$ during Run~2 (2015--2018) ~\cite{DAPR-2021-01}.

%
%
%


%
%
%
%


%
%
%
%
%
%
%
%
%
%

%

%
Monte Carlo (MC) simulation was used to model the detector response to the signal, and to estimate the background contributions.
MC signal samples corresponding to the phenomenological type-I seesaw model were simulated following the prescription given in Ref.~\cite{Fuks:2020att}.
The samples were produced using the \MGNLO[3.5.7]~\cite{Alwall:2014hca} generator at next-to-leading-order (NLO) in QCD,
using the heavy Majorana neutrino Universal FeynRules Output model~\cite{Christensen:2009jx, Alloul:2013bka}.
Parton showering and hadronization were performed with \PYTHIA[8.308]~\cite{Sjostrand:2014zea}, using the A14 set of tuned parameters~\cite{ATL-PHYS-PUB-2014-021}.
The NNPDF3.1~NLO parton distribution function (PDF) set~\cite{Ball:2014uwa} was employed for matrix-element generation.
Signal samples were generated separately for the $\tau\tau$, $e\tau$, and $\mu\tau$ final states, with twenty masses in the range of $m_N = 50$~\GeV to $25$~\TeV,
assuming $|V_{\tau N}|^{2}=1$ and $|V_{eN}|^{2}=|V_{\mu N}|^{2}=0$ for the $\tau\tau$ signal, $|V_{\tau N}|^{2}=|V_{eN}|^{2}=1$ and $|V_{\mu N}|^{2}=0$ for the $e\tau$ signal, and $|V_{\tau N}|^{2}=|V_{\mu N}|^{2}=1$ and $|V_{eN}|^{2}=0$ for the $\mu\tau$ signal.

Processes including the diboson ($VV$), the single $Z$ or $W$ boson in association with jets ($V+$jets), and top-quark pair ($t\bar{t}$) production are simulated to estimate the background contributions.
The $VV$~\cite{ATL-PHYS-PUB-2017-005} and $V+$jets~\cite{PMGR-2021-01} backgrounds are simulated using~\SHERPA[2.2.14]~\cite{Bothmann:2019yzt},
while the \ttbar production~\cite{ATL-PHYS-PUB-2018-009, ATL-PHYS-PUB-2023-029} is simulated using~\POWHEGBOX~\cite{Frixione:2007nw, Nason:2004rx, Frixione:2007vw,Alioli:2010xd}.
The detector response is modeled with a software framework~\cite{SOFT-2010-01} based on \GEANT~\cite{Agostinelli:2002hh},
including the effects of multiple proton--proton collisions per bunch crossing (pileup)~\cite{ATL-PHYS-PUB-2016-017}.


%

%

Collision vertices are reconstructed from at least two ID tracks, each with a transverse momentum \linebreak $\pt > 500$~\MeV~\cite{ATL-PHYS-PUB-2015-026}.
Among all vertices, the one with the highest $\pt^2$ sum of all associated tracks is chosen to be the primary vertex (PV) of the event.
Electrons are reconstructed from ID tracks~\cite{EGAM-2018-01} that are matched to energy clusters in the electromagnetic calorimeter and which originate from the PV\@. The latter condition is satisfied by a requirement on the
transverse impact parameter significance, $|d_0|/\sigma(d_0) < 5.0$, and on the longitudinal impact parameter, $|z_0\sin(\theta)| < 0.5$~mm. They are required to have $|\eta| < 2.47$.
The transition region between the barrel and endcap calorimeters ($1.37 < |\eta| < 1.52$) is excluded.
Muon candidates are reconstructed in the range of $|\eta| < 2.5$ by matching tracks
found in the muon spectrometer to tracks found in the ID~\cite{PERF-2015-10}. Muons reconstructed as stand-alone tracks in the muon spectrometer are also considered.
Like electrons, they have to satisfy impact-parameter requirements: $|d_0|/\sigma(d_0) < 3.0$ and $|z_0\sin(\theta)| < 0.5$~mm. The electrons and muons are required to have a transverse momentum $\pt > 10~\GeV$.
Electrons are required to satisfy \textit{TightLH} identification and \textit{Tight\_VarRad} isolation criteria~\cite{EGAM-2018-01, EGAM-2021-01}.
Muons are required to satisfy the \textit{Medium} identification and \textit{PflowTight\_VarRad} isolation criteria~\cite{MUON-2018-03}.

\par
Jets are reconstructed by clustering particle flow objects~\cite{PERF-2015-09} using the anti-$k_T$ algorithm~\cite{Cacciari:2008gp,Fastjet}
with a radius parameter of $R=0.4$. Jet cleaning criteria are used to remove jets arising from non-collision backgrounds and calorimeter noise.
To suppress central jets from pileup interactions, jets are required to satisfy the jet vertex tagger~\cite{PERF-2014-03} selection if they have $\pT<60$~\GeV and $|\eta|<2.5$. Forward jets are required to satisfy
the baseline of the forward jet vertex tagger~\cite{ATL-PHYS-PUB-2019-026}.
The jets are required to have $\pT>25~$\GeV and $|\eta|<4.5$. Jets originating from $b$-hadrons ($b$-jets) are identified using the GN2 multivariate algorithm, based on a graph neural network, at a working point corresponding to 85\% $b$-jet efficiency~\cite{FTAG-2023-05}.

\par
Hadronic $\tau$ decays are composed of a neutrino and a set of visible decay products~(\tauhadvis),
typically one or three charged pions and up to two neutral pions.
The \tauhadvis candidates reconstructed from seeding jets~\cite{ATL-PHYS-PUB-2022-044} must have $\pt > 20~\GeV$,
$|\eta| < 2.5$ excluding $1.37 < |\eta| < 1.52$, one or three associated tracks and an electric charge of $\pm 1$.
A recurrent neural-network identification algorithm~\cite{ATL-PHYS-PUB-2022-044}, based on calorimetric shower shapes and
tracking information, is used to reject hadronic jets misidentified as $\tau_\text{had-vis}$.
The \tauhadvis candidates must satisfy the \textit{medium} identification criteria~\cite{ATL-PHYS-PUB-2022-044} with efficiencies of about 75\%~(60\%)
for one-track~(three-track) \tauhadvis candidates, respectively. The rejection factors for multijet events are about 30~(500) for one-track~(three-track) \tauhadvis{} candidates.
A separate recurrent neural-network based score (eVeto)~\cite{ATL-PHYS-PUB-2022-044} is used to reject electrons misidentified as $\tau_\text{had-vis}$.

Since the objects used in the analysis are reconstructed from the same set of tracks and
calorimetric energy clusters, some constituents can be associated with multiple objects. These
overlaps between different objects are resolved with the procedure described in Ref.~\cite{HIGG-2019-09}.

\par
The missing transverse momentum, \met, is calculated as the negative vectorial sum of the \pt
of all reconstructed and calibrated physics objects~\cite{JETM-2020-03}.
In addition, this procedure includes a soft term, which is calculated using the ID tracks that
originate from the hard-scattering vertex but are not associated with reconstructed objects.
%


To target the signal with a VBS topology, all events are required to have at least two jets with $\pt > 30~\GeV$, with the two leading jets satisfying $m_{jj} > 300~\GeV$ and $|\Delta \eta_{jj}| > 4$.
Events with $b$-jets are discarded to reject backgrounds with top quarks, with small impact on the signal, which is dominated by light-quark jets.

Events are classified into three signal regions (SRs): the $e\tau_{\text{had}}$ SR and $\mu\tau_{\text{had}}$ SR, which are selected using the $\TauLepHad$ selection described below applied to events with a same-sign electron+$\tauhadvis$ or muon+$\tauhadvis$ pair, respectively,
and the $\tau_{\text{had}}\tau_{\text{had}}$ SR, which targets a same-sign pair of $\tauhadvis$ candidates.

Events in the $\TauLepHad$ SRs are required to fire single-electron or single-muon triggers, with \pt thresholds as low as 24~\GeV and 20~\GeV, respectively~\cite{TRIG-2018-01,TRIG-2018-05,TRIG-2016-01}.
The events must contain at least one \tauhadvis candidate with $\pt$ larger than 55~\GeV and exactly one isolated lepton~($\ell$) with $\pt$ larger than 27~\GeV.
The isolated lepton and the \tauhadvis candidate must have the same electric charge.
Events with extra \tauhadvis candidates are vetoed to ensure orthogonality with the $\tau_{\text{had}}\tau_{\text{had}}$ SR.
The signal acceptance times efficiency for a $\tau\tau$ signal\footnote{The product of acceptance and efficiency includes detector acceptance, reconstruction and selection efficiencies, and is calculated relative to the sum of all $\tau$ decay modes.}
is about 1.1\% in $e\tau_{\text{had}}$ SR and 0.8\% in $\mu\tau_{\text{had}}$ SR for $m_{N}=50$~\GeV, increasing to about 3.1\% and 2.1\%, respectively, for $m_N$ greater than 1~\TeV. The higher efficiency in the $e\tau_{\text{had}}$ SR
compared with the $\mu\tau_{\text{had}}$ SR is due to the looser transverse impact parameter significance requirement and the more efficient electron trigger compared with the muon trigger.

In the $\tau_{\text{had}}\tau_{\text{had}}$ SR, events are selected by single- or di-\tauhadvis triggers, requiring either one \tauhadvis candidate with $\pt$ between 80 and 160~\GeV,
or two \tauhadvis candidates with $\pt$ greater than 35~\GeV and 25~\GeV for the leading and subleading candidate, respectively~\cite{TRIG-2016-01,ATLAS-CONF-2017-061}.
Both \tauhadvis candidates are required to have $\pt$ larger than 55~\GeV and have the same electric charge.
Events with electrons or muons are vetoed to ensure orthogonality with the $e\tau_{\text{had}}$ and $\mu\tau_{\text{had}}$ SRs.
The signal acceptance times efficiency is 1.0\% at $m_{N}=50$~\GeV, increasing to 5.4\% and higher for $m_N$ greater than 2~\TeV.
For all three SRs, the selection efficiencies are driven by the requirements on the \tauhadvis $\pt$ threshold and the VBS topology.



%
For both \TauLepHad and \HadHad channels, the dominant backgrounds are events with a jet misidentified as the \tauhadvis candidate (Jet $\to \tauhad$ fake).
These events are mainly multijet or $W/Z$+jets events in which the jets originate from quarks or gluons.
These contributions are estimated by using the data-driven fake factor method~\cite{TAUP-2023-01}.
The fake factor is derived in a control region enriched in $W$+jets events (WCR).
The definition of the WCR is similar to that of the \muHad channel signal region, but without the requirement on $m_{jj}$ and $\Delta \eta_{jj}$.
In addition, the $\mu$ and \tauhadvis must have opposite charges, $\met>20~\GeV$, and $m_\text{T}(\ptvec^\ell,\metvec)=\sqrt{2\pt^\ell\met[1-\cos\Delta\phi(\ptvec^\ell,\metvec)]}>20~\GeV$.
The fake factor is defined as the ratio of the number of \tauhadvis candidates satisfying the \textit{medium} identification requirement
to the number failing to meet the \textit{medium} but satisfying the \textit{loose} identification requirement.
It is parameterized as a function of $\pt$ and $\eta$, and is derived separately for the one-track and three-track \tauhadvis candidates.
The jet fake background is estimated by applying the fake factor to weight events in an Anti-ID region, which is defined
by reversing the \tauhadvis identification requirement used in the signal regions while still satisfying the \textit{loose} identification criteria.
The contamination by real $\tau$-leptons is estimated with MC simulation, and then subtracted from the data event yield.
The signal contamination in WCR is negligible.

The remaining backgrounds are estimated by using MC simulation and grouped according to the \tauhadvis origin. Backgrounds with a lepton misidentified as a \tauhadvis candidate ($\ell\to \tauhad$ fake)
arise predominantly from electrons in the \TauLepHad channel, contributing to 7\% of the total background. Lepton fake backgrounds are negligible in the \HadHad channel.
This $\ell\to \tauhad$ fake background is estimated by using MC simulation with a data-driven correction derived from $Z\rightarrow ee$ events~\cite{ATLAS-CONF-2017-029}.
Backgrounds where the \tauhadvis candidates originate from prompt $\tau$-leptons arise mainly from $VV$, $Z$+jets and $t\bar{t}$ processes (Others);
they contribute less than 8\% (4\%) of the total background in the \TauLepHad (\HadHad) channel.
In the \TauLepHad channel, $VV$ becomes significant at high $\pt$.

To study the modeling of backgrounds, two validation regions (VR) are defined with the same preselection requirements.
The opposite-sign charge (OS) VR is defined by requiring the $\TauLepHad$ or \HadHad to have opposite electric charges and is enriched with real lepton events, especially $Z$+jets.
The AntiVBS VR is defined by inverting the VBS selections to $m_{jj}<300~\GeV$ and $|\Delta \eta_{jj}|<4$ and is enriched with jet-fake events.
The signal contamination in OS VR and AntiVBS VR is negligible.
The data are found to be in agreement with the predicted background in these two validation regions.
Therefore, no further correction or uncertainty is derived from these two regions.

\par
Uncertainties affecting the signal and the backgrounds are considered in the statistical analysis.
These experimental systematic uncertainties include those related to the trigger, reconstruction, and identification of objects in the final state.
These sources include electron and muon energy or momentum scale and resolution~\cite{EGAM-2021-01,MUON-2018-03}, \tauhadvis{} energy scale
and resolution~\cite{ATLAS-CONF-2017-029}, $b$-tagging efficiency~\cite{FTAG-2018-01,FTAG-2019-02,FTAG-2020-08}, $\met$ reconstruction~\cite{JETM-2020-03},
jet energy scale and resolution~\cite{JETM-2018-05}, and the luminosity measurement~\cite{DAPR-2021-01}.
For heavy Majorana neutrino samples, various sources of uncertainty that affect the signal acceptance are considered, such as the impact of varying
the factorization and renormalization scales and uncertainties in the modeling of initial- and final-state radiation,
as well as multiple parton interactions.
The sensitivity of the search is limited by the statistical uncertainty of the signal region and the Anti-ID region for the heavy Majorana neutrinos of $\mN>1~\TeV$.
Three sources of uncertainty are considered for the fake factor: the statistical uncertainty from the data and MC samples in the WCR;
the uncertainty from subtracting the MC-predicted prompt-lepton contribution in the numerator and denominator regions; and the uncertainty due to the difference in jet composition between the control regions and the signal regions,
since QCD multijet events constitute a major source of jet fakes in the signal region, particularly in the \HadHad channel. This last uncertainty is evaluated by inverting the requirements on $\met$ and transverse mass used to define the WCR\@.
A conservative uncertainty of 30\% is applied to the ``Others'' background processes
to account for the cross-section uncertainty~\cite{STDM-2019-26,STDM-2018-32,STDM-2018-17,TOPQ-2020-09} and the impact is found to be negligible.


%
%

%
%
%
%

The search is conducted using the binned \pT distribution of the leading \tauhadvis{}.
A simultaneous profile likelihood fit is performed in the signal regions of the \eHad, \muHad and \HadHad channels to infer the presence of a signal under the background model assumption.
The likelihood function depends on the free-floating signal normalization and a set of nuisance parameters encoding the effect of systematic uncertainties.
The nuisance parameters~\cite{Rolke:2004mj}, implemented as Gaussian constraints, allow variations of the expectations for signal and background
according to the corresponding systematic uncertainties,
while penalizing the likelihood function for any deviation from the nominal expectations according to the specified constraints.
The statistical uncertainties of background predictions, for example, the Anti-ID region uncertainty, are included through bin-by-bin nuisance parameters.
The fit and the subsequent statistical analysis are performed for each signal hypothesis.
The expected relative uncertainties in the signal-strength value from the leading sources of systematic uncertainties are shown in Table~\ref{tab:uncertainties}.
The dominant uncertainties arise from the statistical uncertainties of the signal region and the Anti-ID region.

\begin{figure}[htb]
\subfloat[]{{\includegraphics[width=0.32\textwidth]{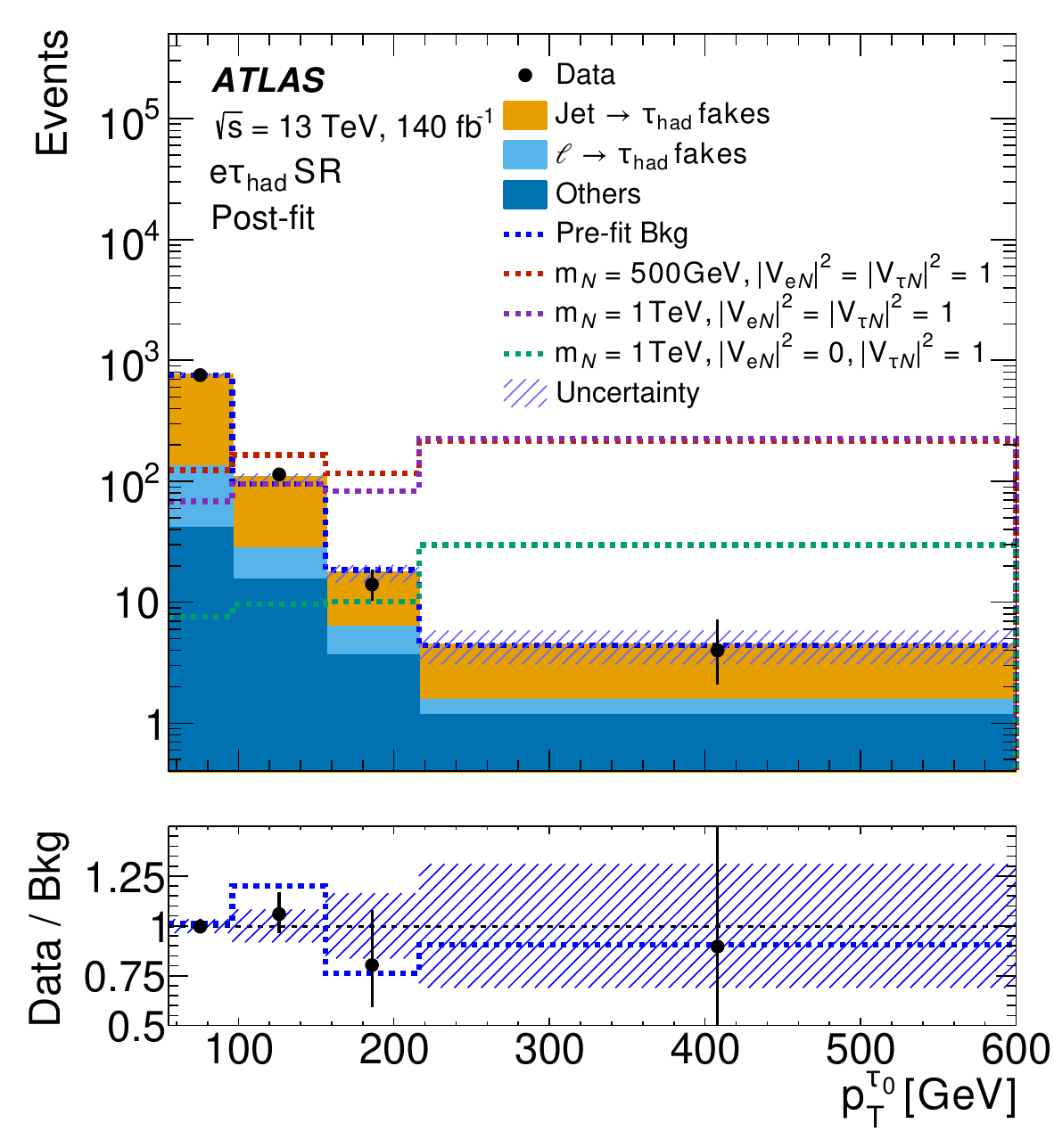}}}\hfill
\subfloat[]{{\includegraphics[width=0.32\textwidth]{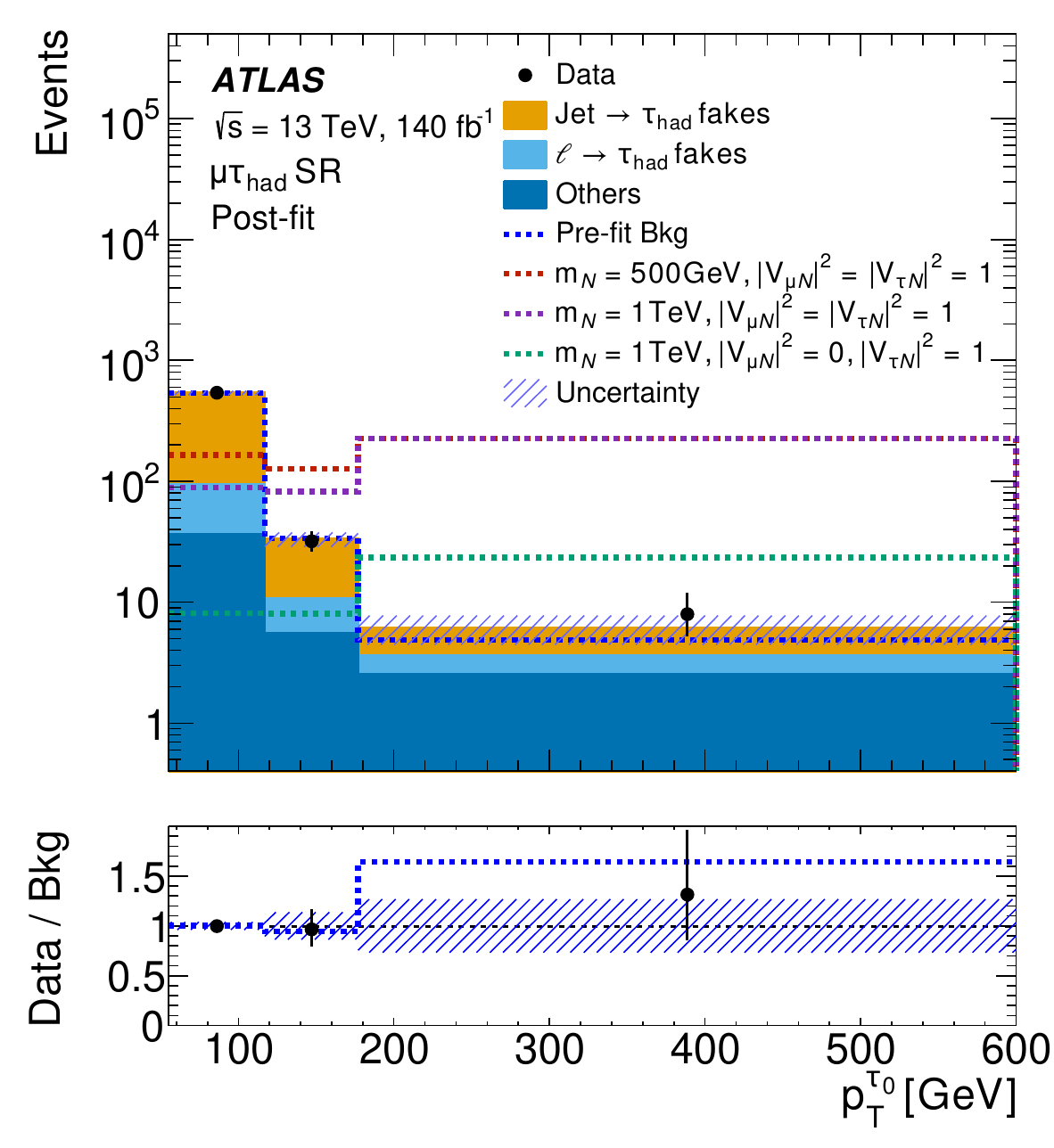}}}\hfill
\subfloat[]{{\includegraphics[width=0.32\textwidth]{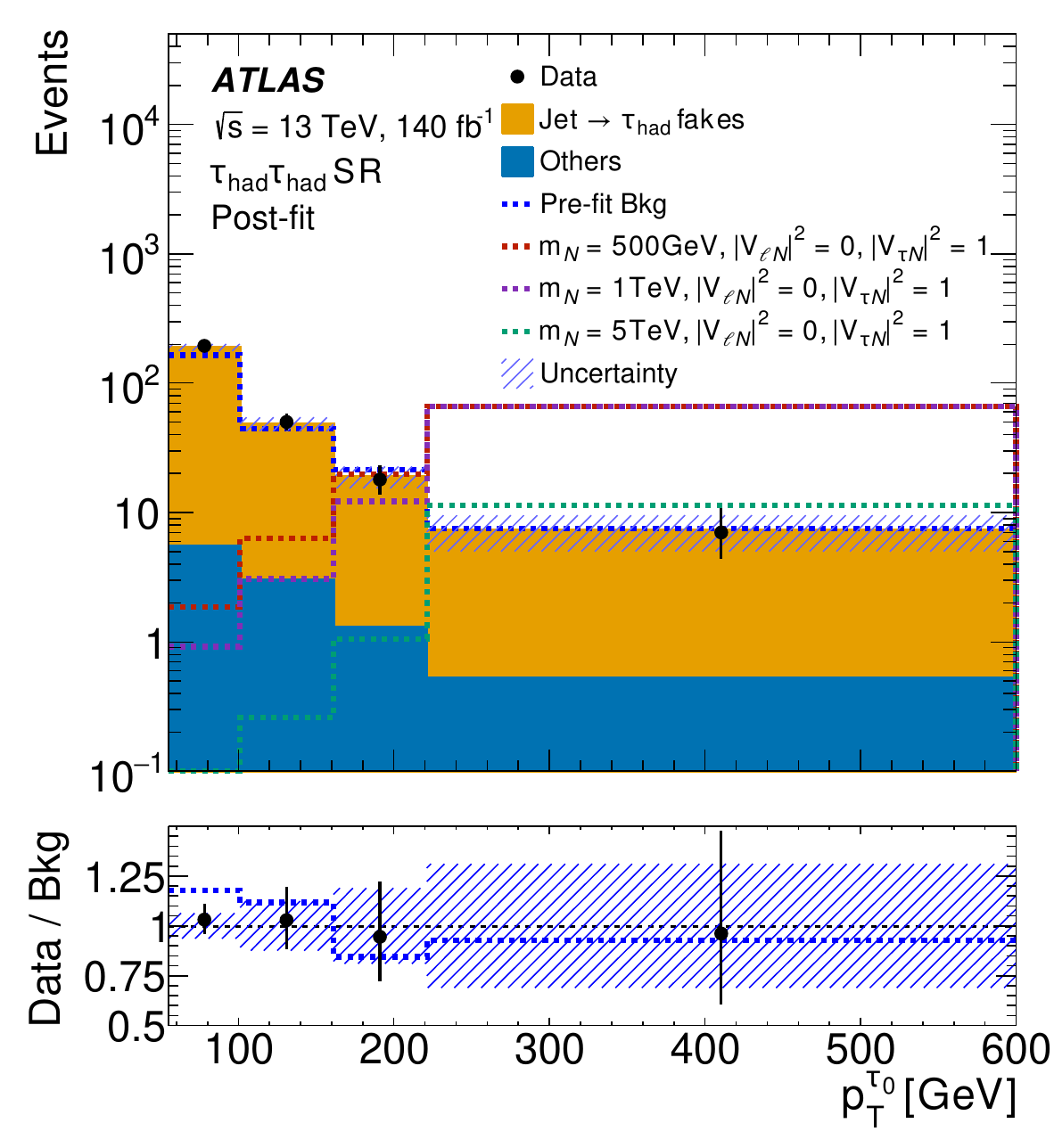}}}
\caption{
The $\pt$ distribution of the leading \tauhadvis{} in the (a) \eHad channel, (b) \muHad channel and the (c) \HadHad channel.
The predictions and uncertainties for the background processes are obtained from the fit assuming the background-only hypothesis.
The expected signal is overlaid for a heavy Majorana neutrino with a mass of 500~\GeV, 1~\TeV and 5~\TeV with different assumptions on the mixing elements;
overflows are included in the last bin of the distributions.
}
\label{fig:pttau_postfit}
\end{figure}

\begin{figure}[htb]
\centering
\includegraphics[width=0.7\textwidth]{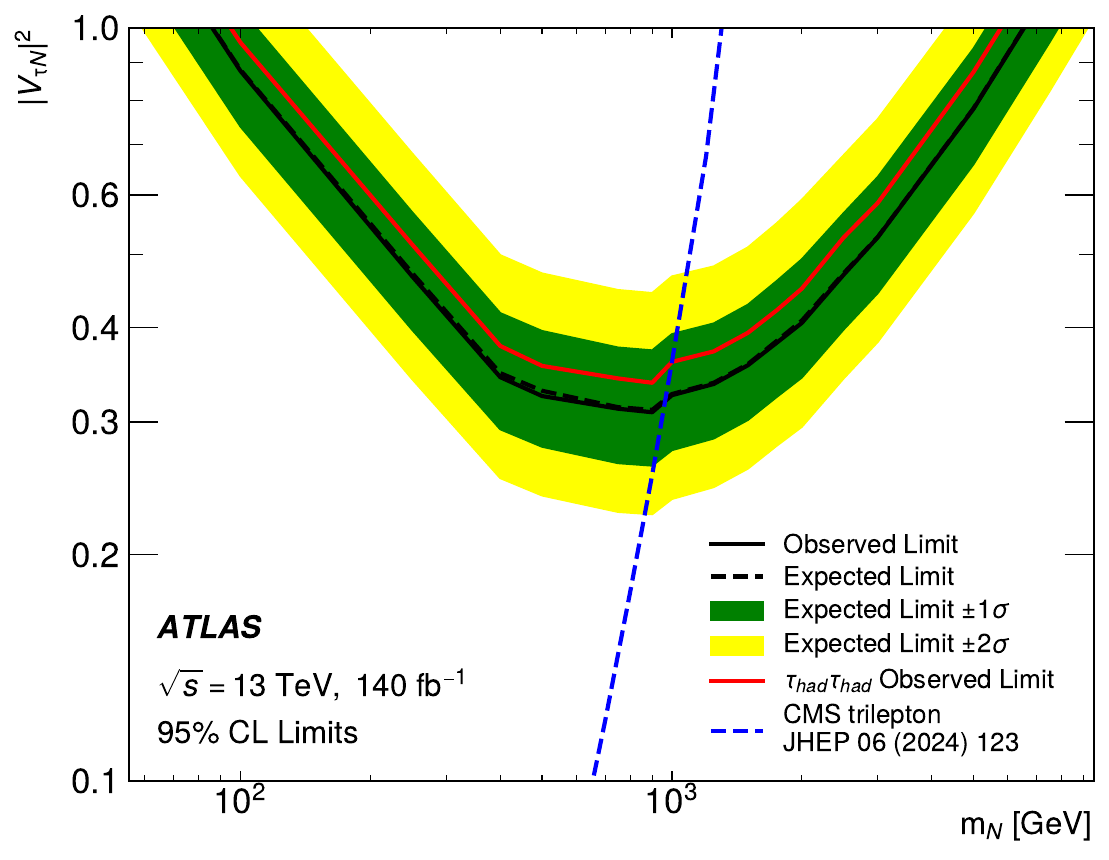}
\caption{
Expected (dashed) and observed (solid) 95\% CL upper limits on the heavy Majorana neutrino mixing element $\abs{ \Vtn }^{2}$
as a function of the heavy Majorana neutrino mass $\mN$, with the assumption of $|V_{e N}|^{2}=|V_{\mu N}|^{2}=0$.
The green and yellow bands represent the $\pm 1 \sigma$ and $\pm 2 \sigma$ uncertainties of the expected limit, respectively.
The red line shows the observed limit from the \HadHad channel.
The feature around 900~\GeV is consistent with a statistical fluctuation found in the reconstruction efficiency of the signal events.
The result from CMS trilepton analysis~\cite{CMS-EXO-22-011} is shown in the dashed blue line from which the region at the left side is excluded.
}
\label{fig:upper_limits}
\end{figure}

\begin{figure}[htb]
\centering
\subfloat[]{\includegraphics[width=0.44\textwidth]{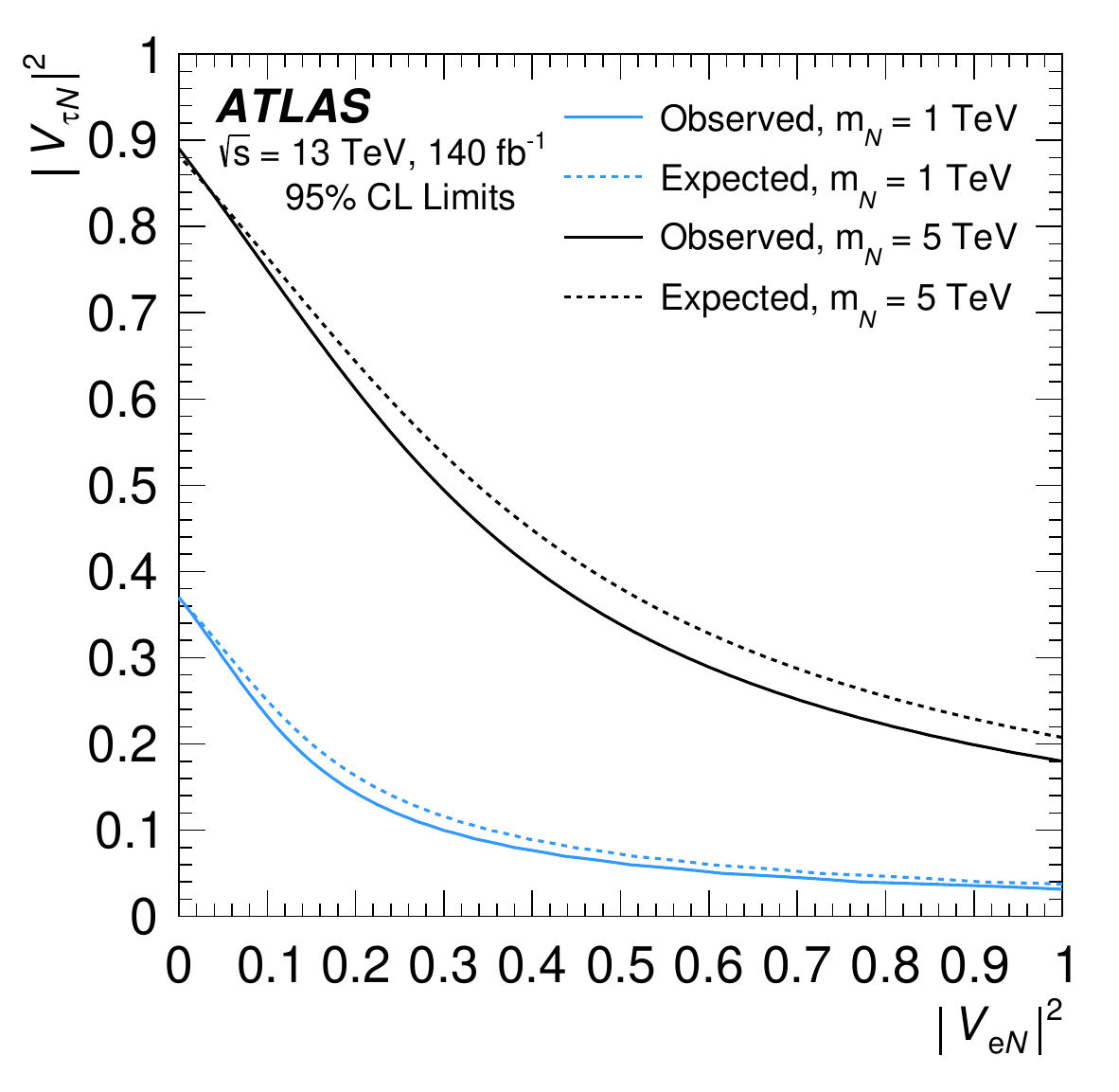}}\hfill
\hspace{0.02\textwidth}
\subfloat[]{\includegraphics[width=0.44\textwidth]{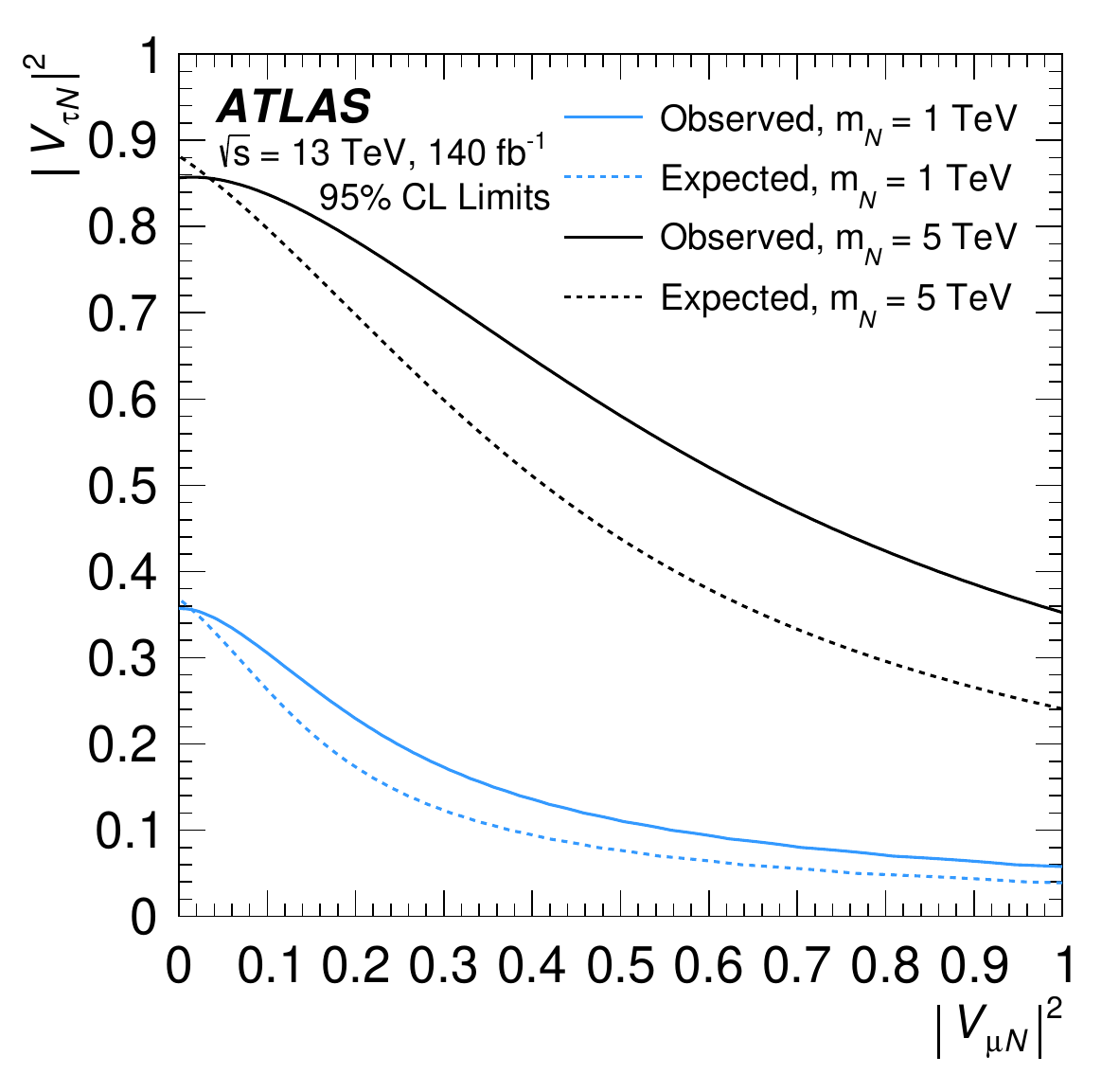}}
\caption{Expected (dashed) and observed (solid) 95\% CL exclusion contours in the (a) $(|V_{eN}|^{2}, |V_{\tau N}|^{2})$
and (b) $(|V_{\mu N}|^{2}, |V_{\tau N}|^{2})$ planes for $m_{N} = 1$~\TeV and 5~\TeV. The area above each curve is excluded.}
\label{fig:2DLimits}
\end{figure}

The leading \tauhadvis{} $\pt$ distributions in the three SRs are shown in Figure~\ref{fig:pttau_postfit}, after performing a fit to the data and assuming a background-only hypothesis. The observed numbers of events in the \eHad, \muHad, and \HadHad SRs are 887, 579, and 270, respectively.
The fitted event yields from background processes are 886$\pm$29, 579$\pm$23 and 264$\pm$14, respectively.
The data are found to be in agreement with the background expectation. Therefore, upper limits at 95\% confidence-level (CL) are derived using
a modified frequentist $\text{CL}_\text{s}$ method~\cite{Read:2002hq} with the profile likelihood ratio as the test statistic.
The asymptotic approximation is used~\cite{Cowan:2010js}.
The mixing elements of the heavy Majorana neutrino can be derived from the cross-section:
$\sigma_{\ell\ell^{\prime}} \propto |V_{\ell N}|^{2} \cdot |V_{\ell^{\prime} N}|^{2}$, where $\ell,\ell^{\prime}= e,\mu,\tau$.
Using this relationship and the upper limits on the cross-section of the heavy Majorana neutrino,
upper limits at 95\% CL on $|V_{\tau N}|^2$ are derived as a function of $m_N$, assuming $|V_{eN}|^{2}=|V_{\mu N}|^{2}=0$, with the observed limit falling below 1 for $m_N$ between $92~\GeV$ and $6.5~\TeV$, as shown in Figure~\ref{fig:upper_limits},
where the \HadHad channel contributes the best sensitivity. %
The feature around 900~\GeV is consistent with a statistical fluctuation found in the reconstruction efficiency of the signal events.
The signal samples with $e\tau$ and $\mu\tau$ final states are also used to constrain two mixing elements simultaneously in the scenarios $(|V_{eN}|^2, |V_{\tau N}|^2, |V_{\mu N}|^2 = 0)$ and $(|V_{\mu N}|^2, |V_{\tau N}|^2, |V_{eN}|^2 = 0)$.
Figure~\ref{fig:2DLimits} presents the results for heavy Majorana neutrino masses $m_N$ of 1~\TeV and 5~\TeV.
It is worth noting that the combination in Ref.~\cite{EXOT-2023-16} excluded at 95\% CL the region $|V_{eN}|^{2}>0.1$ and $|V_{\mu N}|^{2}>0.1$ for $m_N=500$~\GeV, without accounting for $|V_{\tau N}|^{2}$.

\begin{DIFnomarkup}

\begin{table}[htbp]
\caption{The uncertainties (\%) in the signal-strength parameter from each uncertainty source for $\tau\tau$ signal at $m_N=500$~\GeV, 1~\TeV and 5~\TeV.
The first item represents the statistical uncertainty of the signal regions while the second item represents the statistical uncertainty of the Anti-ID regions.}
\label{tab:uncertainties}
\centering
\begin{tabular}{lccc}
\hline
Source & 500~\GeV & 1~\TeV & 5~\TeV \\
\hline
Statistical (signal region) & 3.4 & 3.4 & 20.4 \\
Statistical (Anti-ID region) & 4.1 & 3.9 & 21.5 \\
Fake factor & 1.3 & 1.2 & 6.4 \\
Experimental & 0.8 & 0.8 & 4.1 \\
\hline
Total    & 5.6 & 5.4 & 30.9 \\
\hline
\end{tabular}
\end{table}

\FloatBarrier
\par

\end{DIFnomarkup}

%
%
%

%
%
%
%

In summary, this Letter presents a search for heavy Majorana neutrinos up to several~\TeV in the VBS process using the same-sign $\tau\tau$ final state at the LHC\@.
No significant excess was observed above the expected backgrounds.
The 95\% CL upper limits on the mixing element $|V_{\tau N}|^2$ are set for signals
with heavy Majorana neutrino masses $m_N$ between $92$~\GeV and $6.5$~\TeV, based on a phenomenological type-I seesaw model.
The strongest exclusion was achieved at $m_N = 900~\GeV$, with an observed (expected) limit of 0.30 (0.31).
These results set the most stringent limits on heavy Majorana neutrinos coupling to $\tau$-leptons for $m_N > 1.0~\TeV$
and establish the first direct constraints for $m_N > 1.5~\TeV$.
To date, this represents the highest mass scale explored at colliders for heavy Majorana neutrinos coupling to third-generation leptons.
Furthermore, simultaneous constraints on $(|V_{eN}|^2, |V_{\tau N}|^2)$ and $(|V_{\mu N}|^2, |V_{\tau N}|^2)$ are derived.


%
%
%
%
%
\section*{Acknowledgements}
%

%

%
%

%
%

We thank CERN for the very successful operation of the LHC and its injectors, as well as the support staff at
CERN and at our institutions worldwide without whom ATLAS could not be operated efficiently.

The crucial computing support from all WLCG partners is acknowledged gratefully, in particular from CERN, the ATLAS Tier-1 facilities at TRIUMF/SFU (Canada), NDGF (Denmark, Norway, Sweden), CC-IN2P3 (France), KIT/GridKA (Germany), INFN-CNAF (Italy), NL-T1 (Netherlands), PIC (Spain), RAL (UK) and BNL (USA), the Tier-2 facilities worldwide and large non-WLCG resource providers. Major contributors of computing resources are listed in Ref.~\cite{ATL-SOFT-PUB-2026-001}.

We gratefully acknowledge the support of ANPCyT, Argentina; YerPhI, Armenia; ARC, Australia; BMWFW and FWF, Austria; ANAS, Azerbaijan; CNPq and FAPESP, Brazil; NSERC, NRC and CFI, Canada; CERN; ANID, Chile; CAS, MOST, NSFC and FRFCU, China; Minciencias, Colombia; MEYS CR, Czech Republic; DNRF and DNSRC, Denmark; IN2P3-CNRS and CEA-DRF/IRFU, France; SRNSFG, Georgia; BMFTR, HGF and MPG, Germany; GSRI, Greece; RGC and Hong Kong SAR, China; ICHEP and Academy of Sciences and Humanities, Israel; INFN, Italy; MEXT and JSPS, Japan; CNRST, Morocco; NWO, Netherlands; RCN, Norway; MNiSW, Poland; FCT, Portugal; MNE/IFA, Romania; MSTDI, Serbia; MSSR, Slovakia; ARIS and MVZI, Slovenia; DSI/NRF, South Africa; MICIU/AEI, Spain; SRC and Wallenberg Foundation, Sweden; SERI, SNSF and Cantons of Bern and Geneva, Switzerland; NSTC, Taipei; TENMAK, T\"urkiye; STFC/UKRI, United Kingdom; DOE and NSF, United States of America.

Individual groups and members have received support from BCKDF, CANARIE, CRC and DRAC, Canada; CERN-CZ, FORTE and PRIMUS, Czech Republic; COST, ERC, ERDF, Horizon 2020 and Marie Sk{\l}odowska-Curie Actions, European Union; Investissements d'Avenir Labex, Investissements d'Avenir Idex and ANR, France; DFG and AvH Foundation, Germany; Herakleitos, Thales and Aristeia programmes co-financed by EU-ESF and the Greek NSRF, Greece; BSF-NSF and MINERVA, Israel; NCN and NAWA, Poland; La Caixa Banking Foundation, CERCA and AGAUR programs from Generalitat de Catalunya and PROMETEO and GenT Programmes Generalitat Valenciana, Spain; G\"{o}ran Gustafssons Stiftelse, Sweden; The Royal Society and Leverhulme Trust, United Kingdom; Eric and Wendy Schmidt Fund for Strategic Innovation, United States of America.

In addition, individual members wish to acknowledge support from Chile: Agencia Nacional de Investigaci\'on y Desarrollo (ANID FONDECYT reg. 1230987, FONDECYT 1230812, FONDECYT 1240864, Fondecyt 3240661, Fondecyt Regular 1240721); China: Fundamental Research Funds for the Central Universities (010-63263105), Chinese Ministry of Science and Technology (MOST-2023YFA1605700, MOST-2023YFA1609300), National Natural Science Foundation of China (NSFC 12275265, NSFC-W2543005); Czech Republic: Czech Science Foundation (GACR - 24-11373S), Ministry of Education Youth and Sports (ERC-CZ-LL2327, FORTE CZ.02.01.01/00/22\_008/0004632); EU: H2020 European Research Council (ERC - 101002463); European Union: European Research Council (BARD No. 101116429, ERC - 101219398, ERC 101089007), European Regional Development Fund (HE COFUND GA No.101081355, ERDF), Marie Sklodowska-Curie Actions (GAP-101168829); France: Agence Nationale de la Recherche (ANR-22-EDIR-0002, ANR-24-CE31-0504-01); Germany: Deutsche Forschungsgemeinschaft (DFG - 469666862); China: Research Grants Council (GRF); Italy: Istituto Nazionale di Fisica Nucleare (LHC-MIUR - 28003/2025), Ministero dell'Università e della Ricerca (NextGenEU I53D23000820006 M4C2.1.1, SOE2024\_0000023); Japan: Japan Society for the Promotion of Science (JSPS KAKENHI  JP25H0063, JSPS KAKENHI JP22H04944, JSPS KAKENHI JP24K23939, JSPS KAKENHI JP24KK0251, JSPS KAKENHI JP25H00650, JSPS KAKENHI JP25H01291, JSPS KAKENHI JP25K01011, JSPS KAKENHI JP25K01023, JSPS KAKENHI JP25KK0047); Poland: Polish National Science Centre (NCN 2021/42/E/ST2/00350, NCN OPUS 2023/51/B/ST2/02507, NCN OPUS nr 2022/47/B/ST2/03059, NCN UMO-2019/34/E/ST2/00393, UMO-2022/47/O/ST2/00148, UMO-2023/49/B/ST2/04085, UMO-2023/51/B/ST2/00920, UMO-2024/53/N/ST2/00869); Spain: Agència de Gestió d'Ajuts Universitaris i de Recerca. (AGAUR - 2023 BP 00141), Ministry of Science and Innovation (RYC2019-028510-I, RYC2020-030254-I, RYC2021-031273-I, RYC2022-038164-I), Ministerio de Ciencia, Innovación y Universidades/Agencia Estatal de Investigaci\'on (EU NextGenerationEU (PRTR-C17.I1), PID2022-142604OB-C22); Sweden: Carl Trygger Foundation (Carl Trygger Foundation CTS 22:2312), Swedish Research Council (Swedish Research Council 2023-04654, VR 2021-03651, VR 2022-03845, VR 2022-04683, VR 2023-03403, VR 2024-05451, VR 2025-05940), Knut and Alice Wallenberg Foundation (KAW 2023.0366); United Kingdom: The Binks Trust, Royal Society (NIF-R1-231091); United States of America: U.S. Department of Energy (ECA DE-AC02-76SF00515), John Templeton Foundation (John Templeton Foundation 63206), Neubauer Family Foundation.

%
%


%
\clearpage
\printbibliography
\clearpage
 
\begin{flushleft}
\hypersetup{urlcolor=black}
{\Large The ATLAS Collaboration}

\bigskip

\AtlasOrcid[0000-0002-6665-4934]{G.~Aad}$^\textrm{\scriptsize 102}$,
\AtlasOrcid[0000-0001-7616-1554]{E.~Aakvaag}$^\textrm{\scriptsize 17}$,
\AtlasOrcid[0000-0002-5888-2734]{B.~Abbott}$^\textrm{\scriptsize 121}$,
\AtlasOrcid[0000-0002-0287-5869]{S.~Abdelhameed}$^\textrm{\scriptsize 83b}$,
\AtlasOrcid[0000-0002-1002-1652]{K.~Abeling}$^\textrm{\scriptsize 54}$,
\AtlasOrcid[0000-0001-5763-2760]{N.J.~Abicht}$^\textrm{\scriptsize 48}$,
\AtlasOrcid[0000-0002-8496-9294]{S.H.~Abidi}$^\textrm{\scriptsize 30}$,
\AtlasOrcid[0009-0003-6578-220X]{M.~Aboelela}$^\textrm{\scriptsize 44}$,
\AtlasOrcid[0000-0002-9987-2292]{A.~Aboulhorma}$^\textrm{\scriptsize 36e}$,
\AtlasOrcid[0000-0001-5329-6640]{H.~Abramowicz}$^\textrm{\scriptsize 154}$
% ...
\bigskip
\\

% ...
$^{17}$Department for Physics and Technology, University of Bergen, Bergen; Norway.\\
% ...
$^{30}$Physics Department, Brookhaven National Laboratory, Upton NY; United States of America.\\
% ...
$^{36}$$^{(a)}$Facult\'e des Sciences Ain Chock, Universit\'e Hassan II de Casablanca;$^{(b)}$Facult\'{e} des Sciences, Universit\'{e} Ibn-Tofail, K\'{e}nitra;$^{(c)}$Facult\'e des Sciences Semlalia, Universit\'e Cadi Ayyad, LPHEA-Marrakech;$^{(d)}$LPMR, Facult\'e des Sciences, Universit\'e Mohamed Premier, Oujda;$^{(e)}$Facult\'e des sciences, Universit\'e Mohammed V, Rabat;$^{(f)}$Institute of Applied Physics, Mohammed VI Polytechnic University, Ben Guerir; Morocco.\\
% ...
$^{44}$Physics Department, Southern Methodist University, Dallas TX; United States of America.\\
% ...
$^{48}$Fakult\"{a}t Physik, Technische Universit{\"a}t Dortmund, Dortmund; Germany.\\
% ...
$^{54}$II. Physikalisches Institut, Georg-August-Universit\"{a}t G\"ottingen, G\"ottingen; Germany.\\
% ...
$^{83}$$^{(a)}$Khalifa University of Science and Technology, Abu Dhabi;$^{(b)}$New York University Abu Dhabi, Abu Dhabi;$^{(c)}$United Arab Emirates University, Al Ain;$^{(d)}$University of Sharjah, Sharjah; United Arab Emirates.\\
% ...
$^{102}$CPPM, Aix-Marseille Universit\'e, CNRS/IN2P3, Marseille; France.\\
% ...
$^{121}$Homer L. Dodge Department of Physics and Astronomy, University of Oklahoma, Norman OK; United States of America.\\
% ...
$^{154}$Raymond and Beverly Sackler School of Physics and Astronomy, Tel Aviv University, Tel Aviv; Israel.\\
% ...
$^{a}$ Also at Affiliated with an institute formerly covered by a cooperation agreement with CERN.\\
$^{b}$ Also at An-Najah National University, Nablus; Palestine.\\
$^{c}$ Also at Borough of Manhattan Community College, City University of New York, New York NY; United States of America.\\
$^{d}$ Also at Center for Interdisciplinary Research and Innovation (CIRI-AUTH), Thessaloniki; Greece.\\
$^{e}$ Also at Centre of Physics of the Universities of Minho and Porto (CF-UM-UP); Portugal.\\
$^{f}$ Also at CERN, Geneva; Switzerland.\\
% ...

\end{flushleft}


%

%
%

%

%
%
%
%

\end{document}